\documentstyle[11pt,psfig]{article}
\def\rs{\rm s}
\def\rs1{\rm s^{-1}}

\def\rcm{\rm cm}
\def\rcm2{\rm cm^{-2}}

\def\c2r{\chi^2_\nu}
\def\zeta{A_{Fe}}

\def\etal{{\it et al. }}

\begin{document}
\vspace{1.0cm}
{\Large \bf Broad band spectral properties of Seyfert 1 
galaxies observed with  BeppoSAX }

\vspace{1.0cm}

L. Piro$^{1}$, A. De Rosa$^{1}$, M. Dadina$^{1,2}$,  F. Nicastro$^{1,3}$,
G. Matt$^{4}$, G.C. Perola$^{4}$, P. Grandi$^{1}$, F. Fiore$^{3,5}$, 
T. Mineo$^{6}$, L. Maraschi$^{7}$, F. Haardt$^{8}$

\vspace{1.0cm}
$^1${\it Istituto di Astrofisica Spaziale (IAS), C.N.R., 
         via Fosso del Cavaliere , I-00133 Rome, Italy}\\
    \centerline{piro@ias.fra.cnr.it }

$^2${\it BeppoSAX Science Operation Center, c/o Telespazio, via Corcolle 19, 
I-00131 Rome, Italy}

$^3${\it Osservatorio Astronomico di Roma, Via dell'Osservatorio, I-00144, 
Monteporzio Catone, Italy}

$^4${\it Dipartimento di Fisica "E. Amaldi", Universit\`a degli Studi "Roma Tre", Via della Vasca Navale 84, I-00146, Rome, Italy}

$^5${\it BeppoSAX Science Data Center, c/o Telespazio, via Corcolle 19, 
I-00131 Rome, Italy}

$^6${\it Istituto di Fisica Cosmica ed Applicazioni dell'Informatica CNR, Via Ugo La Malfa 
153, I-90146 Palermo, Italy}

$^7${\it Osservatorio Astronomico di Brera, Via Brera 28, I-20121 Milano, Italy}

$^8${\it Dipartimento di Scienze, Universit\`a dell'Insubria/Polo di Como, Italy} 

\vspace{0.5cm}

%%%%%%%%%%%%%%%%%%%%
\section*{ABSTRACT}
%%%%%%%%%%%%%%%%%%%%
We will present some results on the broad--band observations
of BeppoSAX
of the bright Seyfert galaxies NGC 4151 and NGC 5548.

%%%%%%%%%%%%%%%%%%%%%%%
\section{INTRODUCTION}
%%%%%%%%%%%%%%%%%%%%%%%
In the last ten years the increased sensitivity, 
resolution and bandpass of X-ray missions have drastically 
changed our view of  the X-ray spectrum
of emission line AGN. We have
moved 
from an almost featureless power law into
a complex shape, where several  broad and narrow features,
produced in different sites around the central engine,
are imprinted onto the power law. 
These components span a wide range of energies, sometimes overlapping
with each other.  
An unambiguous determination of each component is then difficult, unless 
{\it simultaneous} broad--band spectral measurements are secured.
BeppoSAX, with its 0.1-200 keV range, 
appears particularly suited to undertake a broad-band study
of AGN in X-rays.
In this contribution we will present some results of  Core
Program ( and Science Verification Phase) observations
devoted to the study of Broad-band spectral variability of Seyfert 1 
galaxies.
The scientific goals of the program are:
\begin{itemize}
\item Probe  the environment near the central source. 
This  is achieved by disentangling each spectral component and studying
 its temporal behaviour, in particular the response to 
changes of the intrinsic continuum and the  correlation with other
components. 
\item Investigate the origin of the intrinsic continuum. The key spectral
parameters of the continuum, i.e. the slope $\alpha$ and the
high energy cut-off $E_c$ can be determined with an unprecedented 
precision by BeppoSAX. This should allow to investigate the 
presence of correlation between those parameters and the luminosity, 
an important test-point for radiative models.
\end{itemize}

The baseline observing strategy we chose
is  that of long looks of those bright Seyfert 1
galaxies characterized by   variability time-scale of $\sim$ day. 
This
assures a contiguous  sequence of spectra 
with the S/N needed for 
spectral measurements up to $\sim 100 keV$ and without
substantial variation within each time
bin ($\le $ variability time--scale).

We will present here some of the results obtained during the Science Verification Phase of NGC~4151, and during the AO1 on NGC~5548.

\begin{figure}
\centerline{\psfig{figure=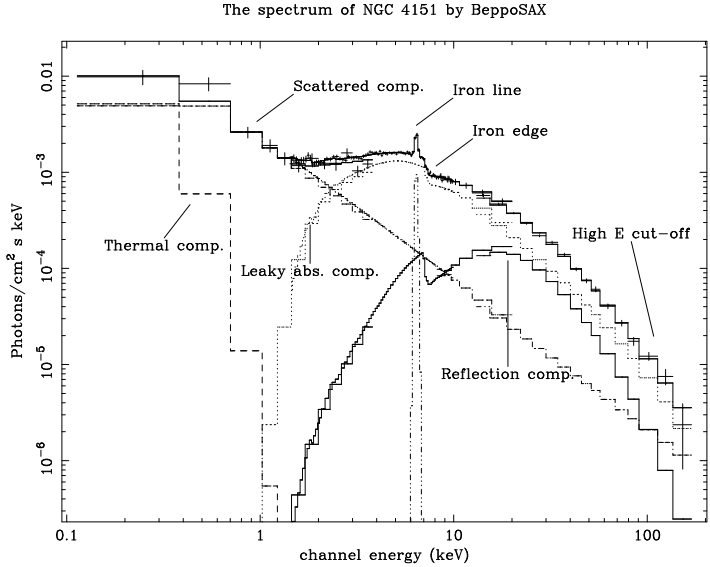,height=8cm,angle=-90}}
\caption{The BeppoSAX spectrum of NGC 4151}
\end{figure}

\begin{figure}
\centerline{\psfig{figure=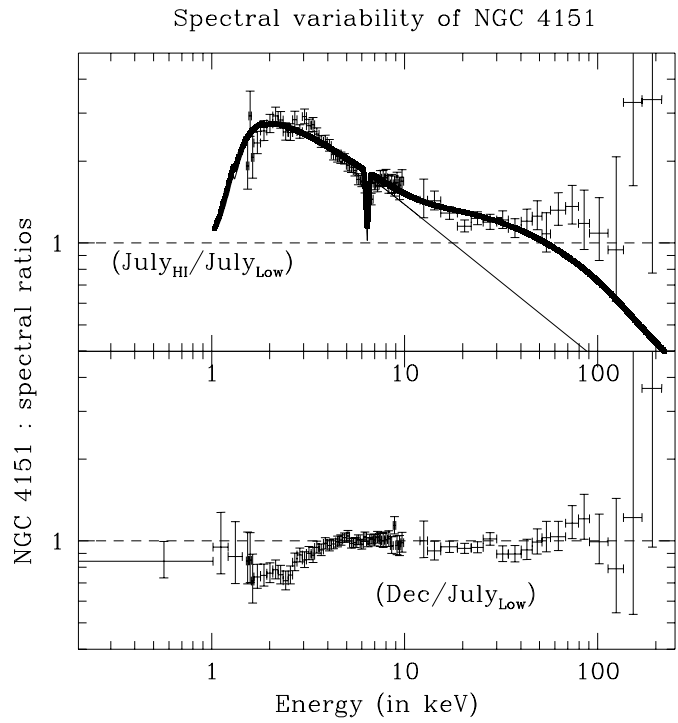,height=10cm,angle=0}}
%\centerline{\psfig{figure=ratio41.ps,height=10cm,angle=0}}

\caption{Spectral variability from BeppoSAX observations
of NGC 4151. Upper panel: ratio of spectra of July 96 High and Low (taken 
2 days apart). Lower panel: Dec. 96 over July 96 Low.
}
\end{figure}

\begin{figure}
%\centerline{\psfig{figure=counts_ratio_55.ps,height=10cm,angle=0}}
\centerline{\psfig{figure=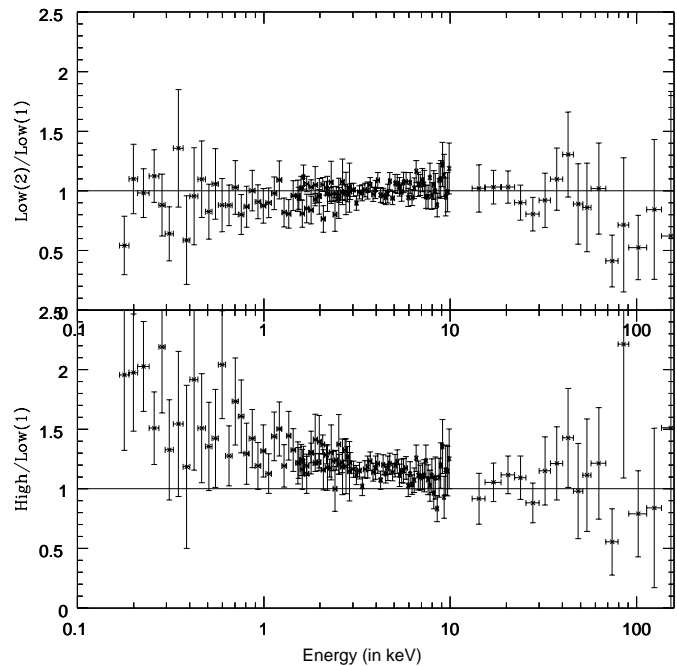,height=10cm,angle=0}}
\caption{Spectral variability of of  NGC~5548 from
the BeppoSAX long--look} 
\end{figure}

\begin{figure}
\centerline{\psfig{figure=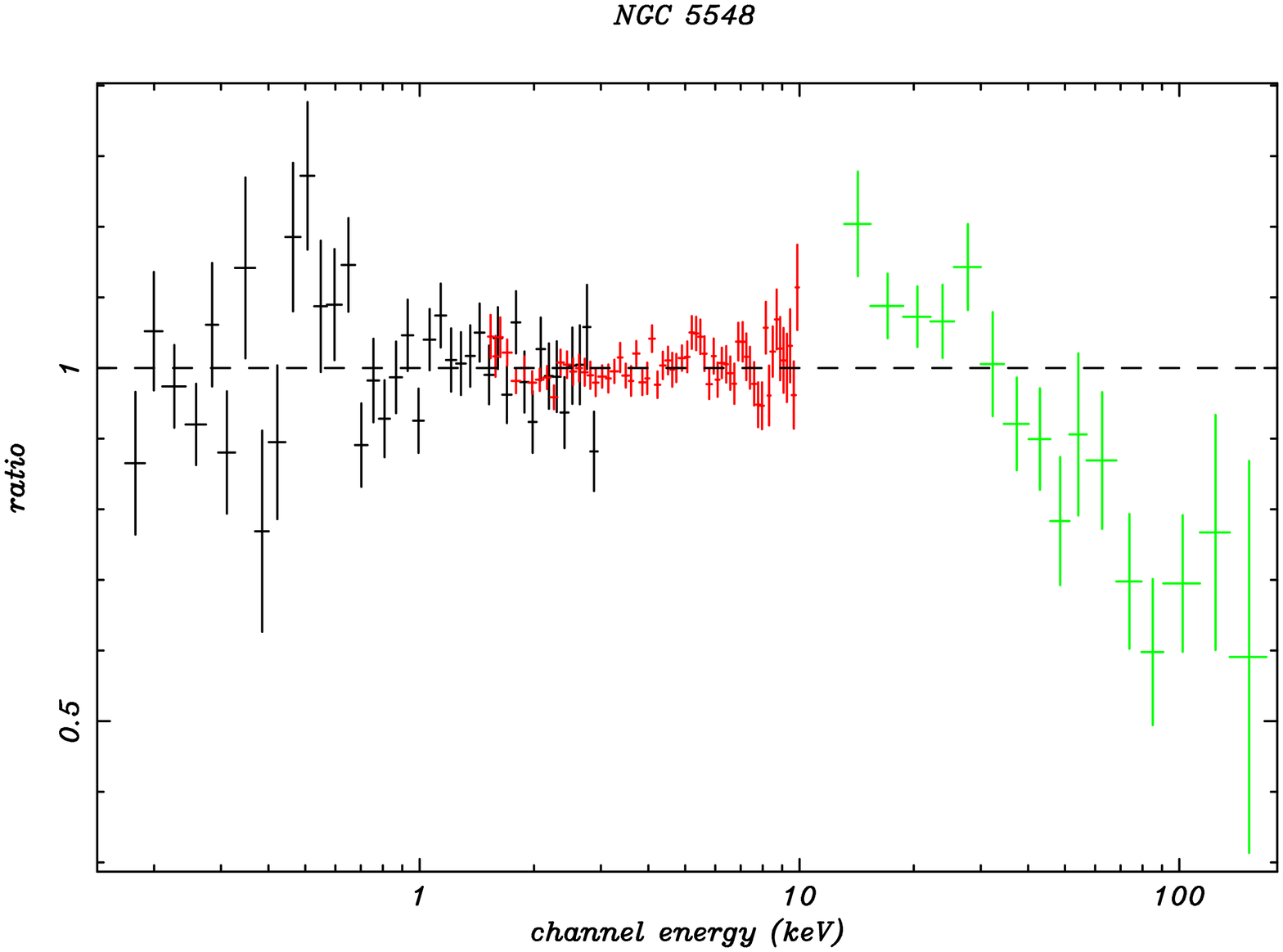,height=8cm,angle=-90}}
%\centerline{\psfig{figure=res55nc.ps,height=8cm,angle=-90}}
\caption{The complex broad band spectrum of NGC~5548 as 
seen by BeppoSAX. The residuals from the best model (before BeppoSAX)
plotted in figure
show the presence of an high energy cut-off and an emission line
around 0.6 keV}
\end{figure}

%----------------------------------figure 1-----------------------------------

\section{Broad--band spectral variability: 
a probe of the environment and of 
the origin of the intrinsic continuum}

The X-ray spectrum of 
NGC4151 is the most complex 
ever observed in a Seyfert galaxy, as the broad-band picture
of BeppoSAX clearly shows in Fig.1.
The spectrum is characterized by features typical of both 
Seyfert 1 and Seyfert 2 galaxies, making NGC 4151 the 
best laboratory for the study of these objects.
The different temporal behaviour of these components yields
the complex spectral variability showed by this object.
In fig.2 we present the ratio of spectra taken 2 days apart
($July_{HI}, July_{Low}$) and few months apart ($Dec$) in 1996.
Let us first comment the long term behaviour (lower panel of Fig.2).
All the variations can be attributed to a change in the structure
 of the absorber.
The intrinsic continuum (cfr the ratio above 3-4 keV) remained
unchanged, while below 1 keV the predominant 
constant soft components (Perola \etal 1986, Weaver \etal 1994)
force to 1 the spectral ratio.
The variability observed in July on $\sim$ day time--scale
has a different origin.
The factor--of--two flux increase is associated with a
steepening of the intrinsic power law ($\Delta \alpha\sim 0.3$),
fully consistent with the $\alpha\ $vs. $F_X$ relationship
observed first with EXOSAT by
Perola \etal 1986 and then confirmed with GINGA by Yaqoob \etal 1993.
Fig.2 (upper panel) shows that the 2-10 keV spectral variability
is indeed well reproduced by a change of the intrinsic slope.
The presence of
constant soft components dumps down to 1 the spectral ratio
below 2 keV, but above 10 keV the
ratio should continue to decrease with energy (thin line), 
contrary to what
observed. This is fixed  with 
a constant reflection component, whose presence 
is already required by spectral fitting (Piro \etal 1998),
(thick line).
Note also as the intensity of the iron line remained constant,
notwithstanding the substantial change of the ionizing flux
(see also Perola \etal 1986).
This suggests that the line region is
 far from the central source and
 possibly is the
same site of the reflection component.

Intrinsic spectral variability narrows substantially the
range of 
models of
 the intrinsic continuum.
This point has not received much attention by theoreticians
(with some noticeable exceptions), probably
because  the only solid evidence of intrinsic spectral
variability  was that of NGC 4151.
With the broad-band spectral capability of BeppoSAX we
can address this issue with much less ambiguities than in the
past.
Indeed, in the long-look observation of NGC 5548 we find a behaviour similar
to that observed in NGC 4151 
(fig. 3).

\section{The high energy cut-off}

With the exception of NGC 4151, where the high energy cut-off $E_C$
is fairly well determined (Piro \etal 1998 and references therein),
in other Seyfert galaxies only an average value has been derived
(Zdziarski \etal 1995).
With BeppoSAX we can measure $E_C$ in single
objects, with the perspective of studying its correlation
with  $\alpha$ and
the luminosity ( or the compactness parameter).
The long observation of NGC 5548 has allowed to determine
fairly precisely the cut-off ($E_C=140_{-30}^{+60}$ keV, Nicastro
\etal 1999), even with the source at a rather low flux (Fig.4).
It should be also possible, at least in the brightest objects,
to search for changes of $E_C$ correlated with intensity.
Indeed, in NGC 4151, there may be an indication of a variation
of $E_C$
from the low to the high state (upper panel of Fig.3).

%----------------------------------figure 1-----------------------------------
%-------------------------table-1--------------------------------------------

%%---------------------------------
%%\subsection{\underline{qq}}
%%---------------------------------

\section{REFERENCES}
\vspace{-5mm}
\begin{itemize}
\setlength{\itemindent}{-8mm}
\setlength{\itemsep}{-1mm}

%\item[]
%Boella, G., Butler, R., Perola, G., 
%et al., {\it A\&AS}, {\bf 122}, 299 
%(1997a).   

%\item[]

\item[]
Nicastro F., Piro L., De Rosa A. \etal 1999, ApJ, submitted
\item[]
Perola G.C., Piro L., Altamore A. \etal, 1986, ApJ, 306, 508
\item[]
Piro L., Nicastro F., Feroci M. \etal 1998, Nucl. Phys. B, 69/1-3, 481
\item[]
Weaver K.A., Yaqoob T., Holt S. \etal 1994, ApJ, L27
\item[] 
Yaqoob T., Warwick R., Makino F. \etal 1993, MNRAS, 262, 435
\item[] 
Zdziarski A., Johnson N., Done C. \etal 1995, ApJ, 438, L63 
%\item[]

\end{itemize}

\end{document}